\documentclass[12pt]{article} 

\usepackage{times} 
\usepackage{wrapfig,lipsum,booktabs,amssymb,amsmath}
\usepackage{setspace}
\usepackage{bm} 
\usepackage{color} 
\usepackage{hyperref} 
\usepackage[font=small, belowskip=-14pt,aboveskip=0pt]{caption}
\usepackage{graphicx} 
\usepackage{vmargin} 
\usepackage{titlesec}
\usepackage{multirow}
\usepackage[normalem]{ulem}
\usepackage{braket}
\usepackage{cancel}
\usepackage[style=nature,citestyle=numeric-comp, sorting=none, backend=biber, maxbibnames=9]{biblatex}
\usepackage{makecell}
\titlespacing{\section}{5pt}{\parskip}{-\parskip}
\titlespacing{\subsection}{3pt}{\parskip}{-\parskip}
\titlespacing{\subsubsection}{2pt}{\parskip}{-\parskip}

\hypersetup{colorlinks=true, citecolor=black,linkcolor=blue,urlcolor=blue}
\DeclareCaptionFont{blue}{\color{blue}}
\usepackage{etoolbox}
\patchcmd{\thebibliography}{\section*{\refname}}{}{}{}
\usepackage{authblk}

\setpapersize{USletter}

\setmarginsrb{1in}{1in}{1in}{1in}{0pt}{0mm}{0pt}{0mm}

\newcommand{\pseudodot}{{\lower 2.4pt\hbox{$\cdot$}}}

\begin{document}
\setlength{\baselineskip}{12.6pt} 
\setlength{\normalbaselineskip}{12.6pt} 
\title{Twisted Superconducting Quantum Diodes for High Fidelity Anharmonic Qubits} 
\date{\vspace{-5ex}}

\author[1]{Han Zhong}
\affil{Department of Electrical and Computer Engineering, University of Florida, Gainesville, FL 32611, USA}

\author[2,3]{Denis Kochan}
\affil[2]{Department of Physics and Center for Quantum Frontiers of Research and Technology (QFort), National Cheng Kung University, Tainan 70101, Taiwan}
\affil[3]{Institute of Physics, Slovak Academy of Sciences, 84511 Bratislava, Slovakia}
\author[4]{Igor \v{Z}uti\'c}
\affil[4]{Department of Physics, State University of New York at Buffalo, Buffalo, New York 14260, USA}
\author[1,*]{Yingying Wu}
\affil[*]{Corresponding author: yingyingwu@ufl.edu}
\maketitle
\begin{abstract}
In superconducting circuits, bidirectional signal propagation can cause back-reflections and back-action that couple environmental noise back to coherent elements, degrading fidelity. Nonreciprocal superconducting diodes provide an intrinsic route to suppress such back-scattering at the circuit level. However, their diode efficiency and quantum integration remain limited. Here, we realize a quantum diode in twisted NbSe$_2$ under 
in-plane and out-of-plane magnetic fields. A mere 1$^\circ$ twist yields an efficiency enhancement over pristine devices, reaching 27.6\%. Quantum simulations reveal that this intermediate efficiency—well below 100\% ideal—is both experimentally practical and
optimal for preserving qubit anharmonicity and stabilizing two-level systems. These findings show that maximal rectification is not always beneficial for quantum information, establishing a new principle for designing 
the fundamental properties of twisted superconductors towards 
low-power, high-fidelity quantum circuits.
\end{abstract}

\newpage
\section*{Introduction}
Superconducting diodes unify the dissipationless properties of superconductivity with the unidirectional current flow of traditional semiconductor diodes~\cite{nadeem2023superconducting,Amundsen2024:RMP}.
Similar to how semiconductor 
diodes exhibit nonreciprocal resistive charge transport, superconducting diodes demonstrate nonreciprocity in the superconducting state, allowing superconducting currents to flow preferentially in one direction while being suppressed in the opposite direction. This behavior arises from a combination of symmetry-breaking mechanisms, such as spin-orbit coupling, applied magnetic fields, 
$\mathbf{B}$, or intrinsic material properties~\cite{Amundsen2024:RMP}.
By integrating key superconducting characteristics-such as zero electrical resistance, low noise, and high coherence~\cite{Tafuri:2019}
with the directional control of current flow, superconducting diodes hold transformative potential for 
energy-efficient computing architectures, to enable superconducting logic and memory elements with minimal power dissipation, as well as support ultrafast operation, including THz applications~\cite{golod2022demonstration,Monroe2024:APL,yang2023terahertz}.
Furthermore, superconducting diodes offer exciting prospects for 
quantum sensing and ultra-sensitive magnetometry~\cite{Tafuri:2019}, by exploiting their nonreciprocal superconducting transport 
properties~\cite{nadeem2023superconducting}. 

Recent studies show that the superconducting diode effect can be realized in both Josephson junctions and junction-free superconductors~\cite{nadeem2023superconducting,Hou2023ubiquitous}.
Related advances are usually characterized 
by the efficiency of a superconducting diode which is defined as the difference between the critical currents, $I_{\rm C}$, in the forward, and reverse directions, normalized by 
their sum~\cite{nadeem2023superconducting,Pekerten2024:APL,CuzzoPRR:2024}
\begin{equation}
\eta = \frac{|I_{\rm C}^+| - |I_{\rm C}^-|}{ |I_{\rm C}^+| + |I_{\rm C}^-|} \times 100\%,
\label{eq:eta}
\end{equation}  
where $I_{\rm C}^+$ ($I_{\rm C}^-$) denotes the critical current in the forward  (reverse) direction. 
The form of Eq.~(\ref{eq:eta}) conveniently quantifies the asymmetry between 
$I_{\rm C}^+$ and $I_{\rm C}^-$, with many works aiming to maximize $\eta$ and the corresponding rectification~\cite{nadeem2023superconducting,Amundsen2024:RMP}.
The push towards a high-diode efficiency is further motivated by the prospect that in van der Waals (vdW) heterostructures their atomically sharp interfaces could lead to high {\em on}/{\em off} ratios. 
Experimental
diode efficiency in vdW heterostructures~\cite{lin2022zero} is comparable 
to the state-of-the art values $\eta\sim 70$\% in sputtered materials and their heterostructures~\cite{Hou2023ubiquitous,golod2022demonstration}. An additional advantage of vdW heterostructures based on high-temperature superconductors~\cite{zhao2023time}
is that a large $\eta\sim 50$\% can be maintained at higher operating temperature than in superconducting diodes from sputtered materials.  
To date, the superconducting diodes have been proposed primarily for their use as current rectifiers \cite{nadeem2023superconducting,ingla2025efficient}, which explains the primary focus on enhancing the corresponding $\eta$. A similar
trend in $\eta$ can be also seen from the concept of a quantum diode 
designed for the integration with of solid-state qubits and characterized by transport nonreciprocity, enabling unidirectional excitation transfer between the boundary resonators of a lattice. 
The quality of the quantum diode is characterized by its fidelity, which can be improved by increasing $\eta$, 
when the one-way excitation transfer process becomes more
pronounced~\cite{zhao2023engineering,yang2026skyrmion}. 

However, how the diode efficiency influences 
much broader functionalities remains poorly understood, but it is essential to unlock their full potential. In quantum circuits the relevant objective can instead be maximizing fidelity/anharmonicity, subject to noise and design constraints.
Surprisingly, our work on vdW twisted superconducting quantum diodes reveals that an {\em intermediate}, experimentally realized, and not ideal efficiency of $\eta \rightarrow 100$\% can enable their optimal performance. Such twisted systems 
offer a 
platform to study fundamental phenomena and the interplay between time-reversal symmetry breaking, topology, spin-orbit coupling, and strongly-correlated 
effects~\cite{lin2022zero,zhao2023time,Cao2018:N,wang2020topological}.

Here we explore the twisted layers of vdW superconductor NbSe$_2$ and report 
$\eta=27.6$\%.  
A twist angle of just $1^{\circ}$, enhances $\eta$ nearly an order of magnitude 
from a pristine NbSe$_2$~\cite{zhang2024intrinsic}, 
while the spontaneously broken time-reversal symmetry can lead to a diode effect even under nominal near-zero-field conditions, ($\mathbf{B} \approx 0~\mathrm{T}$). We note that a residual magnetic field of several Oe due to trapped flux in the superconducting magnet cannot be completely excluded and may contribute to the observed field offset.
Guided by the measured $\eta$, we simulate a scalable quantum diode design that provides strong anharmonicity for qubit operations and enables seamless integration into quantum  
circuits. Our approach elucidates the role of $\eta$ on
quantum functionalities and reexamines the integration 
of superconducting diodes into a growing landscape of quantum architectures.

\section*{Results}
\subsection*{Diode design }

NbSe$_2$ is a transition metal dichalcogenide which, with a weak interlayer vdW bonding, allows its exfoliation into thin flakes or even monolayers. 
A strong intrinsic spin-orbit coupling of a monolayer NbSe$_2$ pins the spins
of electrons with opposite momenta to opposite out-of-plane direction. 
In the superconducting state such a spin-momentum locking favors Ising 
pairing~\cite{wickramaratne2020ising,wu2019induced,han2018investigation}, 
the formation of singlet-like Cooper pairs with electrons spins pinned perpendicular to the monolayer, and protects the pairs against in-plane spin
polarization, withstanding a much large in-plane $\mathbf{B}$. At very small twist angles, lattice relaxation produces extended commensurate stacking domains separated by solitonic boundaries; this reconstruction strongly modifies the interfacial electronic structure and can enhance inversion-symmetry breaking and interfacial spin-orbit 
textures, both of which promote a diode response. For marginally twisted NbSe$_2$, both theory and experiments indicate especially strong reconstruction below $\sim 1.2^\circ$ for certain registries~\cite{mchugh2023moire}. We have fabricated twisted NbSe$_2$ structure using a twist angle of 1$^{\circ}$. Figure~\ref{Fig1}a presents a schematic representation of a twisted NbSe$_2$ system with a twist angle of $\theta$ and Fig.~\ref{Fig1}b is its corresponding electrical characterization setup with four terminals. The temperature-dependent resistance measurement reveals a superconducting transition at a critical temperature of 6.87 K. The inset of Fig.~\ref{Fig1}c presents a false color image of the fabricated NbSe$_2$ device, highlighting different regions and components involved in the electrical characterization.
Yellow and green parts represent, respectively, the gold electrodes and the h-BN encapsulation layer, which is preventing the oxidation and degradation of the NbSe$_2$ layers. The dark layer corresponds to the twisted NbSe$_2$ layers, whose superconducting properties are the focus of the work. Prior studies of superconductivity in twisted multilayers include other vdW materials such as 
graphene~\cite{Cao2018:N,oh2021evidence} and high-temperature superconductor,  Bi$_2$Sr$_2$CaCu$_2$O$_{8+x}$~\cite{zhao2023time}, but NbSe$_2$-based superconducting diodes were largely explored without any twist angle~\cite{nadeem2023superconducting,Amundsen2024:RMP}.

\subsection*{Results on critical currents and diode effect}

\begin{figure}[ht]
\begin{centering}
 \includegraphics[width=0.7\textwidth]{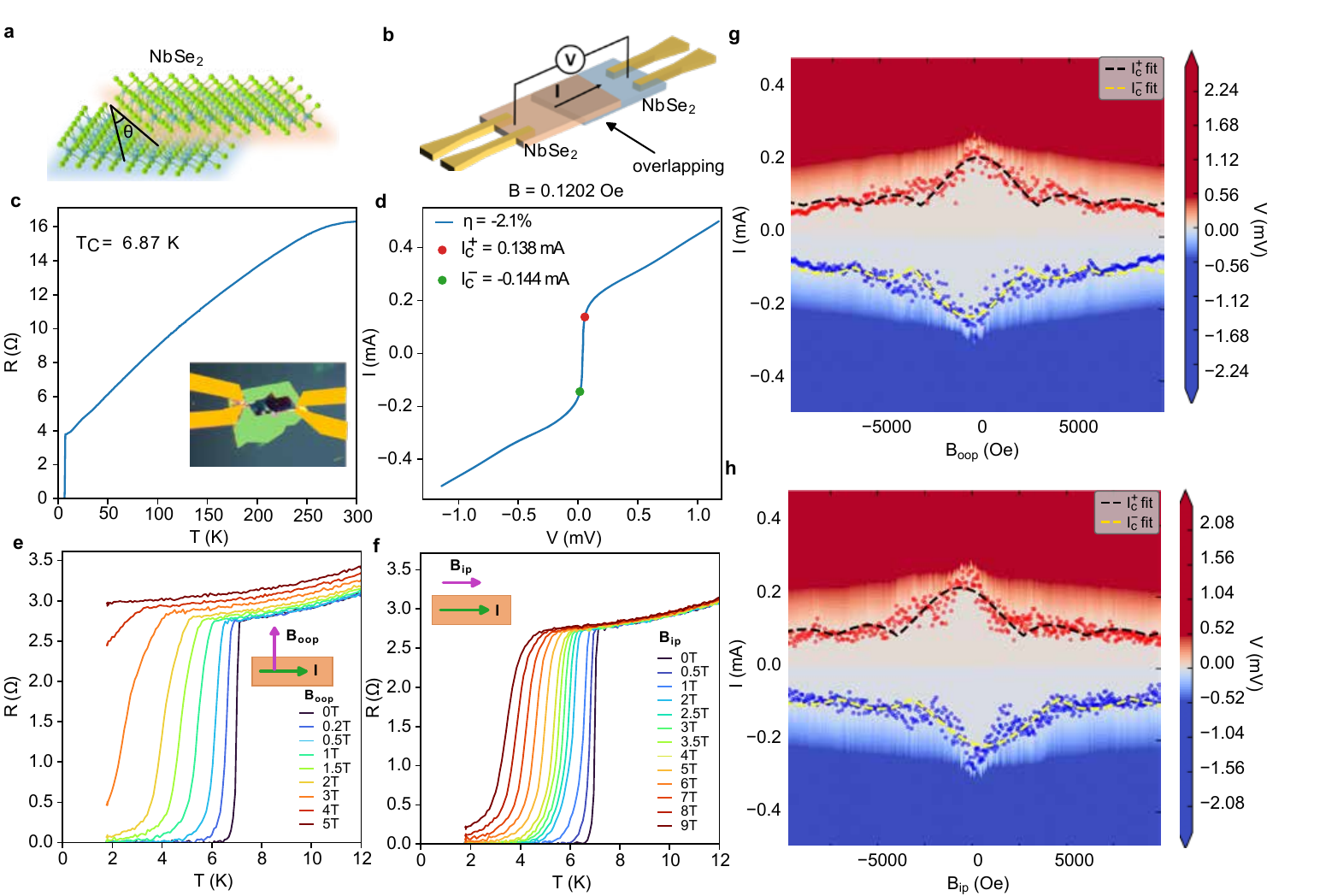} 
 \caption{Diode design of twisted NbSe$_2$ layers. (a) Schematic of twisted layers with an angle $\theta$. (b) Current $I$ and voltage $V$ measurements using four terminal setup. (c) Dependence of the resistance $R$ on temperature $T$ indicating a critical temperature. $T_\mathrm{C}=6.87\,$K, which was extracted from high-resolution $R$-$T$ data. (d) Typical $I$-$V$ characteristics and diode efficiency $\eta$ under an applied in-plane magnetic field close to 0 T and determination of the corresponding critical currents $I_\textrm{C}^+$ and $I_\textrm{C}^-$. (e) Out-of-plane $\mathbf{B}_\text{oop}$ and (f) in-plane magnetic field $\mathbf{B}_\text{ip}$ dependence of $T_\mathrm{C}$.
 All data are from a sample with an overall thickness of 16$\,$nm. (g,h) Voltage maps as a function of 
 $I$ and 
 $\mathbf{B}$, with the extracted 
 $I_\textrm{C}^+$ ($I_\textrm{C}^-$),
 overlaid as red (blue) symbols. The black and yellow dashed lines represent the corresponding Fraunhofer-like fits to $I_\textrm{C}^{+}(B)$ and $I_\textrm{C}^{-}(B)$. }\label{Fig1}\par\end{centering}
\end{figure}

While NbSe$_2$-based superconducting diodes exhibit at $\mathbf{B}=0$ a vanishing rectification efficiency \cite{Amundsen2024:RMP}, our measurements yield $\eta \approx -2.1\%$. This indicates a finite intrinsic anisotropy in the device under
$\mathbf{B} < 10^{-4}$\,T, 
as shown in Fig.~\ref{Fig1}d.
This can be attributed to an emergent correlated phase surviving weak fields that possess a spontaneous twist-induced magnetization stemming from a combination of strong electronic interactions, underlying moir\'{e} landscape and a non-trivial band topology. Their interplay can give rise to a state with orbital magnetization and thus spontaneously broken time-reversal symmetry as reported for nonmagnetic vdW materials with small twist 
angles~\cite{li2021quantum}.
However, applying an out-of-plane ${\mathbf{B}}_\text{oop}$, 
Fig.~\ref{Fig1}e, modulates the resistance, demonstrating the field's influence on superconducting properties. Similarly, resistance measurements under varying in-plane ${\mathbf{B}}_\text{ip}$,
~Fig.~\ref{Fig1}f, displays anisotropic ${\mathbf{B}}_\text{ip}$ response, further emphasizing the role of external magnetic fields in tuning the electronic properties of the twisted NbSe$_2$ system. The enhanced resilience of superconductivity to ${\mathbf{B}}_\text{ip}$ in a few-layer NbSe$_2$, compared to ${\mathbf{B}}_\text{oop}$, is widely attributed to Ising spin-orbit coupling and Ising pairing~\cite{xi2016ising}.
These findings indicate spin-orbit coupling-affected superconductivity~\cite{Amundsen2024:RMP}, possibly an unconventional Ising pairing due to the characteristic vdW spin–momentum locking.
Crucially, to confirm that the
underlying properties are compatible with
Josephson junctions, in Figs.~\ref{Fig1}g-h we  also map current-voltage ($I$-$V$) characteristics
as $\mathbf{B}$ is varied, with the 
evolution of $I_C$ described by
\begin{equation}
I_\textrm{C}(B)=I_{\textrm{bg}}+I_{\textrm{C}0}
\left|
\frac{\sin\!\left[\pi(B-B_0)/B_p\right]}
{\pi(B-B_0)/B_p}
\right|,
\label{eq:fraunhofer}
\end{equation}
where $I_{\mathrm{bg}}$ is a field-independent background contribution, $I_{\mathrm{C}0}$ is the amplitude of the Fraunhofer-like modulation~\cite{Tafuri:2019}, 
$B_0$ is the $\mathbf{B}$ offset 
corresponding to the center of the interference pattern, and $B_p$ is the characteristic field period that determines the width of the central lobe and the spacing of the side lobes. The observed peak in $I_\mathrm{C}$ near $B=0$
and its oscillatory suppression at larger 
$\mathbf{B}$ 
are consistent with Josephson-interference-like behavior~\cite{yabuki2016supercurrent} in the twisted NbSe$_2$ junction.

\begin{figure}[ht]
\begin{centering}
\includegraphics[width=0.9\textwidth]{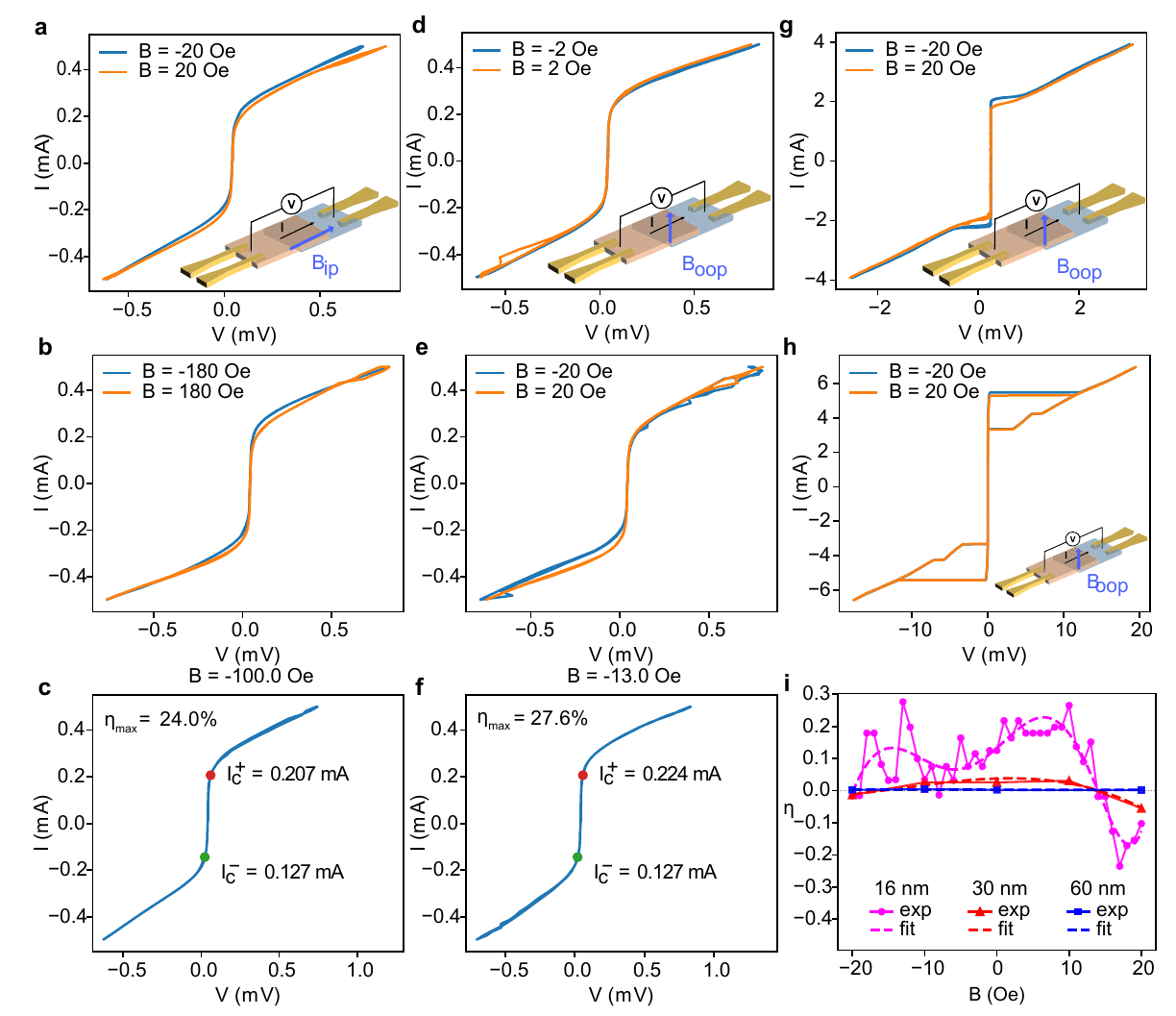} 
 \caption{$I$-$V$ characteristics of the twisted NbSe$_2$ with overall thickness of (a-f) 16\,nm, (g) 30$\,$nm, and (h) 60\,nm. (a) $I$-$V$ curves at $\mathbf{B}_\text{ip}=\mp 20\,$Oe applied along the current. (b) The same as in (a) but at $\mp180\,$Oe.  (c) In-plane maximum $\eta$ is achieved at $\mathbf{B}_\text{ip}=-100\,$Oe. (d) $I$-$V$ curves
 at $\mathbf{B}_\text{oop}=\mp 2\,$Oe. (e) The same as in (d) but at $\mathbf{B}_\text{oop}=\mp 20\,$Oe. (f) The maximum diode efficiency is achieved at $\mathbf{B}_\text{oop}=-13\,$Oe. (g) $I$-$V$ curves at $\mathbf{B}_\text{oop}=\mp 20\,$Oe for overall 30 nm thick sample.
(h) The same as in (g) but for 60 nm thick sample. (i) Diode efficiencies and fittings as functions of $\mathbf{B}_\text{oop}$ for NbSe$_2$ junctions with different overall thicknesses: 16$\,$nm (magenta), 30$\,$nm (orange), and 60$\,$nm (blue), 
see the corresponding color coding. Note $\mathbf{B}=0\,\mathrm{T}$ denotes the nominal applied field; a residual magnetic field of several Oe due to trapped flux in the superconducting magnet cannot be completely excluded.\label{Fig2}}
\par\end{centering}
\end{figure}

After fabricating twisted NbSe$_2$ devices with three representative overall thicknesses--16~nm, 30~nm, and 60~nm (corresponding to 8~nm, 15~nm, and 30~nm per NbSe$_2$ layer)--and performing resistance-temperature (R-T) measurements, we further investigate their $I$-$V$ characteristics under both 
${\mathbf{B}}_\text{ip}$ and ${\mathbf{B}}_\text{oop}$. Figure~\ref{Fig2} shows the $I$-$V$ curves and $\eta$, as ${\mathbf{B}}$ is varied, extracted using a standard procedure~\cite{moll2025geometrical}.
Our measurements demonstrate $\mathbf{B}$-dependent nonreciprocal transport behavior, essential for the superconducting diode effect. 

Figures~\ref{Fig2}a-c, show $I$-$V$ curves for the device with overall thickness of 16$\,$nm $\mathbf{B}_\text{ip}=\pm$20\,Oe and $\pm$180\,Oe applied (anti)parallel to the current with the maximum 
$\eta \approx 24.0$\% at -100\,Oe. These curves, which correspond to the forward
and backward current sweeps, nearly coincide in each of the panels. The asymmetry in the $I$–$V$ curves suggests broken inversion symmetry in the superconducting state, where is is predicted that interfacial spin–orbit coupling can give rise to a superconducting diode effect~\cite{kang2024magnetotransport}. 
In twisted vdW heterostructures, density functional theory calculations indicate that such coupling may include a non-zero radial Rashba component~\cite{veneri2022twist,frank2024emergence}. Correspondingly, Figs.~\ref{Fig2}d-f, display $I$-$V$ curves under $\mathbf{B}_\text{oop}=\pm$2\,Oe and $\pm$20\,Oe with the maximum $\eta \approx$ 27.6\% at -13\,Oe, demonstrating a strong
nonreciprocal response with an only small perpendicular field. This suggests that the system experiences the vortex diode effect~\cite{vodolazov2005superconducting} as also observed in other systems~\cite{sivakov2018spatial,Hou2023ubiquitous,suri2022non} in 
${\mathbf{B}}_\text{oop}$. Twisting the two NbSe$_2$ layers creates an asymmetry between its edges that impacts on different heights of the associated Bean-Livingston surface barriers \cite{bean1964surface}, and hence on different onsets of the resistive states depending on the polarity of current drive, or 
$\mathbf{B}_\text{oop}$
~\cite{ingla2025efficient,castellani2025a,kochan2025two}. 

Figures~\ref{Fig2}g-h show representative $I$-$V$ curves measured at 
$\mathbf{B}_\text{oop} = \pm20\,$Oe for devices with an overall thickness of 30$\,$nm and 60$\,$nm. In both cases, a clear nonreciprocity is observed in the $I$-$V$ responses, with asymmetry in the forward and reverse critical currents. A clear hysteresis behavior shown in 60$\,$nm device in {Fig.~\ref{Fig2}h} 
can be explained using the standard resistively and capacitively shunted
junction picture~\cite{Tafuri:2019}: 
A hysteresis is expected when the damping is weak, quantified by the Stewart--McCumber parameter $\beta_c=(2e/\hbar)I_{\rm C} R^2 C>1$. Increasing the device thickness increases the effective capacitance $C$ and eventually $I_{\rm C}$, both of which
enhance $\beta_c$ and hence the hysteresis, even for clean interfaces. Figure~\ref{Fig2}i summarizes $\eta$ as a function $\mathbf{B}_\text{oop}$ 
for junctions of different thicknesses. The 16$\,$nm device exhibits a pronounced 
$\mathbf{B}$-dependent diode efficiency, while the 30$\,$nm and 60$\,$nm devices show 
a much weaker nonreciprocal behavior highlighting the role of interlayer coupling and symmetry breaking in controlling the rectification effect. The thickness dependence of the nonreciprocity originates from the vortex diode effect: In thicker samples, the Meissner screening current density is reduced and the effective Pearl penetration depth, $\lambda_{\text{Pearl}}=\lambda_{\text{London}}^{2}/d$, becomes smaller, which in turn enhances the asymmetry of the Bean–Livingston vortex surface barriers, 
where $\lambda_{\text{London}}$ is the London penetration depth~\cite{Tafuri:2019} and $d$ is the sample thickness.
Consequently, in thicker samples the vortices are bounded 
to the edges more firmly than in the thinner films, what suppresses transition into the resistive state. These results demonstrate the tunability of the superconducting diode effect in twisted NbSe$_2$ through external magnetic fields, with distinct optimal values for $\mathbf{B}_\text{ip}$ and $\mathbf{B}_\text{oop}$.
This behavior suggests an interplay between moir\'e-induced electronic states, interfacial coupling phenomena, and vortex diode effect, altogether influencing the transport properties of the twisted NbSe$_2$.

\begin{figure}[ht]
\begin{centering}
\includegraphics[width=0.94\textwidth]{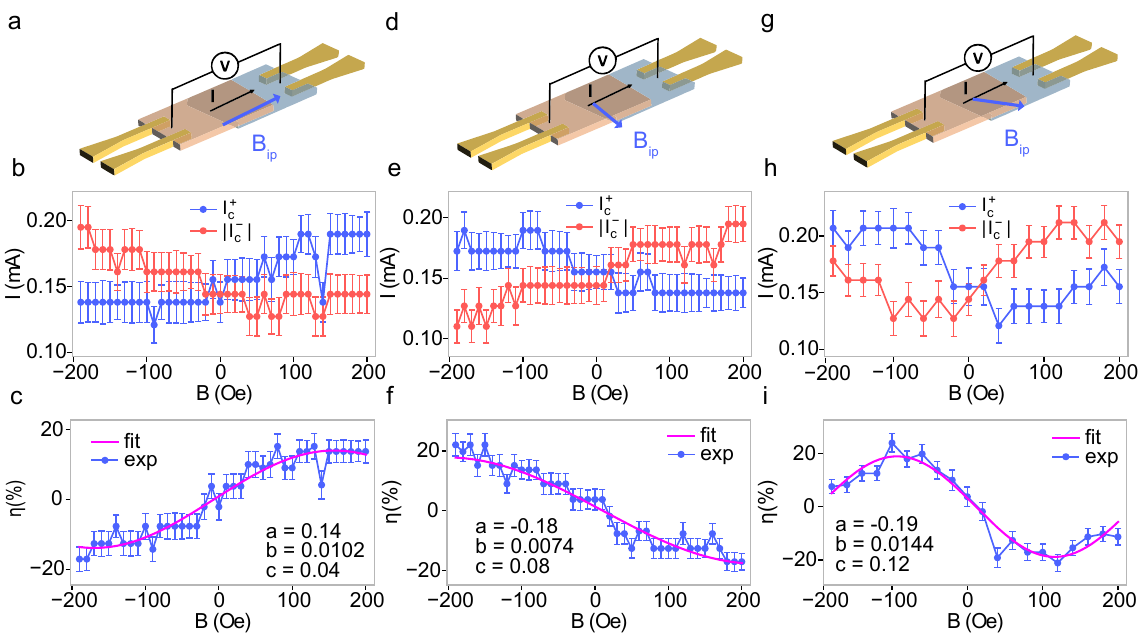} 
 \caption{Diode effect for various $\mathbf{B}_\text{ip}$ configurations 
 measured in a sample with overall thickness of 16$\,$nm. (a)-(c) show 
 $\mathbf{B}_{\rm ip}\backslash\backslash I$, with 
 $I_C$ extracted under varying $\mathbf{B}$ in (b), and $\eta(\mathbf{B})$ in (c). (d)-(f) show the diode characteristics for
${\mathbf{B}_{\rm ip}} \perp I$, with the corresponding $I_C$ in (e) and $\eta$ in (f). (g)-(i) show the same as before for a random in-plane angle between the ${\mathbf{B}}_{\rm ip}$ and $I$,
with $I_C$ in (h) and $\eta$ in (i). (c), (f), and (i) also show the least-squares fitting (fit) of the experimental data (exp) of $\eta$, that is in the range of $\mathbf{B}$-field from -200 to 
 200$\,$Oe following $\eta = a \sin(b \mathbf{B} + c)$, the corresponding values of $a, b$ (in units of 1/Oe) and $c$ are given.
 In (b), (c), (e), (f), (h), and (i) vertical lines denote the calculated uncertainties.}
\label{Fig3}
\par\end{centering}
\end{figure}

The behavior of twisted NbSe$_2$ under different orientations of ${\mathbf{B}}_\text{ip}$ 
reveals distinct dependencies of the diode effect on a mutual orientation of the field and current drive. Figure~\ref{Fig3} shows schematics and experimental results of $I_{\rm C}$ and $\eta$ 
for a device with an overall thickness of 16 nm, exposed to the in-plane magnetic ${\mathbf{B}}$ parallel (Figs.~\ref{Fig3}a-c) and perpendicular (Figs.~\ref{Fig3}d-f) to the current. 
A similar analysis is performed also in the case when angle between the in-plane ${\mathbf{B}}$ and current is arbitrary (Figs.~\ref{Fig3}g-i). In all three cases the corresponding critical currents and diode efficiencies indicate a strong directional dependence of superconducting transport, where the orientation of ${\mathbf{B}}_\text{ip}$ significantly influences the nonreciprocal supercurrent behavior. These results highlight the role of twisting and spin-orbit coupling 
engineering at the interface of two NbSe$_2$ layers that, correspondingly, imparts the anisotropic superconducting diode effect features sensed by ${\mathbf{B}}_\text{ip}$.

A quantitative estimate is added showing that misalignment of the sample with
respect to ${\mathbf{B}}$ cannot explain the magnitude of the in-plane response. 
For 
an angular error $\delta\theta\lesssim 0.1^\circ$, $\mathbf{B}_\text{ip}$=200\,Oe 
would produce an unintended $\mathbf{B}_\text{oop} \simeq \mathbf{B}_\text{ip}\sin\delta\theta\approx 0.35$\,Oe. In contrast, 
$\mathbf{B}_\text{oop}$ scale, where the strongest vortex-diode response occurs, is $\sim 10$-15\,Oe (Fig.~2f). To mimic $\mathbf{B}_\text{oop}\sim$ 13\,Oe at $\mathbf{B}_\text{ip}$=200\,Oe would require $\delta\theta\simeq \arcsin(13/200)\approx 3.7^\circ$, far larger than our alignment tolerance. Therefore, field misalignment alone cannot account for the observed ${\mathbf{B}}_\text{ip}$ dependence.

The error bars for $I_{\rm C}$ were calculated using bootstrap resampling~\cite{bland2015statistics}, that estimates uncertainty without assuming a specific probability distribution. The process involves repeatedly drawing random samples with a replacement from the original $I$-$V$ measurements near the critical point. In the code, this is implemented with 100 resamplings 
for each $I_{\rm C}$, sampling from a window of up to 11 data points (5 on each side of the detected critical point). For each resampled dataset, $I_{\rm C}$ 
is recalculated using the same voltage gradient method as applied to the original data. The standard deviation of these 100 recalculated $I_{\rm C}$ values  
provides a robust estimate of measurement uncertainty, revealing how 
$I_{\rm C}$ determination might vary with slightly different experimental data points.
For the analysis of $\eta$, we use an empirical sinusoidal relationship with an applied 
$\mathbf{B}$, which is modeled as $\eta = a \sin(b \, \mathbf{B} + c)$, where $a$ and $b$ 
parametrize the amplitude and frequency, while $c$ determines the 
phase shift. The fitting algorithm employs a weighted least-squares minimization.
By weighting each data point according to its uncertainty, the fit gives greater influence to more reliable measurements. The resulting high 
coefficient of determination, denoted as $R^2$ (typically \textgreater0.93 in this case), confirms
that this simple first-harmonic parameterization
effectively captures the observed $\mathbf{B}$-dependent diode behavior.

\subsection*{Quantum simulations: A one-way transfer of qubits}

\begin{figure}[ht]
\begin{centering}
 \includegraphics[width=0.95\textwidth]{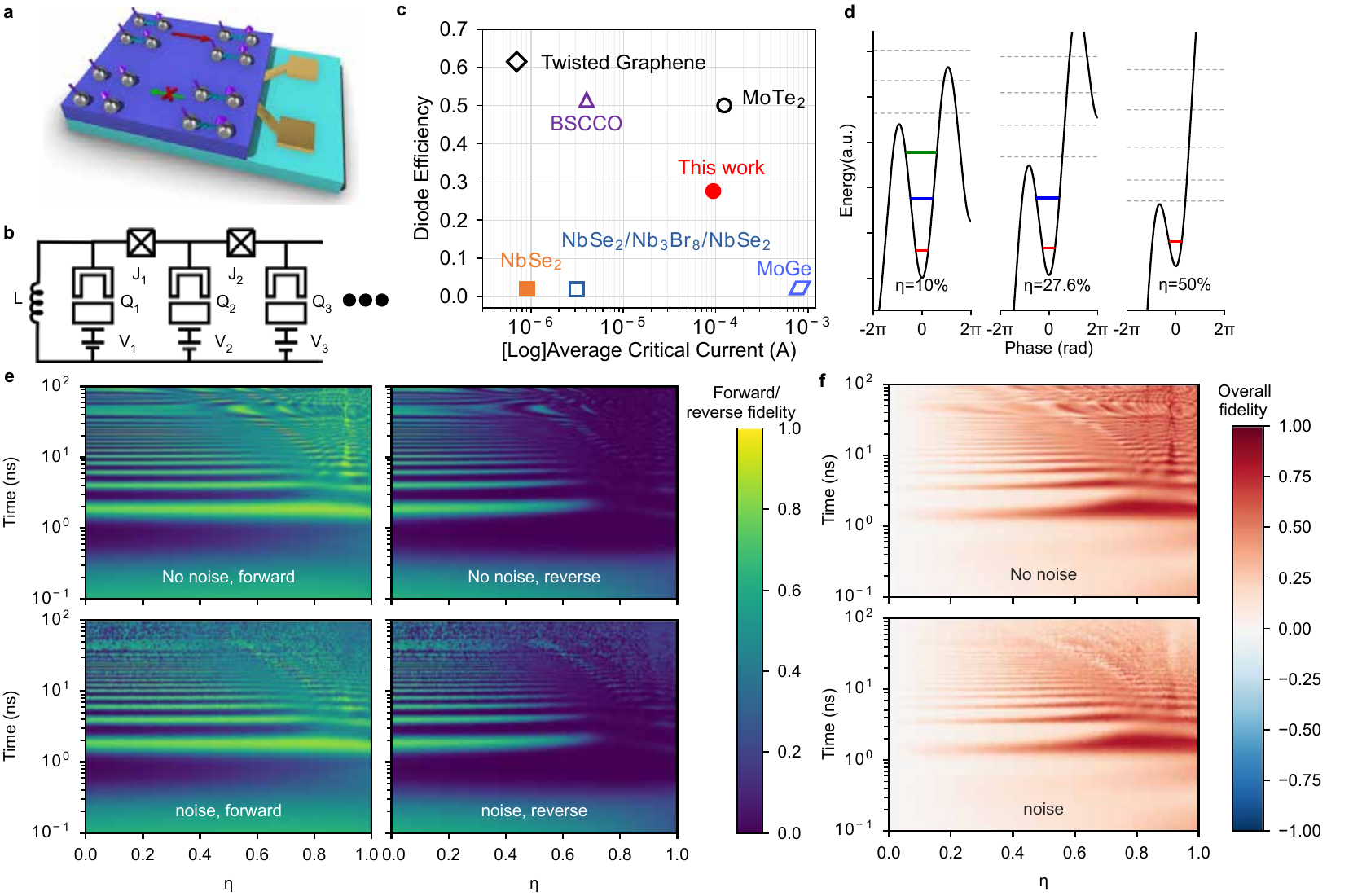} 
 \caption{Quantum circuit design and simulation. (a) Schematics of the NbSe$_2$ device exhibiting a nonreciprocal behavior. (b) Transmon qubit chain circuit with quantum diodes. (c) Benchmark plot comparing diode efficiency and average critical current in reported 2D layered systems, with this work highlighted in red. The efficiencies were calculated from experiments \cite{wu2022field,lee2021twisted,chen2024edelstein,chen2024asymmetric,zhao2023time,lin2022zero,lyu2021superconducting}. (d) Simulation of the potential-energy landscape with the corresponding energies of the qubit levels inside the quantum wells for the three representative efficiencies $\eta$, a perfect two-level system can be achieved for $\eta$ close to 27.6\%. (e) Forward and reverse quantum diode fidelity 
 simulations---as functions of the efficiency and time---without (above) and with (below) noise (based on circuit in (b)). (f) Overall fidelity difference between forward and reverse qubit transfer without (above) and with (below) noise.\label{Fig4}}
\end{centering}
\end{figure}

After characterizing the rectification properties of the twisted NbSe$_2$, we designed superconducting circuits incorporating such nonreciprocal elements, and simulated their quantum diode characteristics. 
Leveraging a widely used, gate-based simulation framework of QuTip, an open-source Python framework for simulating open quantum system dynamics~\cite{johansson2012qutip}, we integrated diode efficiency
$\eta$ as a new degree of freedom that is capable to tune the nonreciprocity of the
 the quantum diode. 
Figure~\ref{Fig4}a shows a schematics of nonreciprocal circuit based on the NbSe$_2$ bilayer junction: Cooper pairs flow with minimal dissipation in the forward direction, while they break into quasiparticles for the reverse bias and thus turning a circuit into a resistive state. 
Figure~\ref{Fig4}b depicts particular platform utilizing transmon qubits ($Q_1, Q_2, Q_3$) interconnected by the NbSe$_2$-based Josephson junction $J_1$, $J_2$ (marked by ``$\boxtimes$" symbols). This provides both, the nonlinearity that is essential for the transmon qubits~\cite{Tafuri:2019}, but also an intrinsic supercurrent nonreciprocity. Such built-in directionality not only preferentially 
steers supercurrents, but also suppresses backward excitations, thereby enhancing a qubit control. 
($Q_1, Q_2, Q_3$) are individually voltage-biased ($V_1, V_2, V_3$) and share a common inductor on the left, enforcing unidirectional quantum‐information flow, crucial for isolating signals and preventing backscattering in a scalable and 
highly-coherent 
way.
Additionally, Fig.~\ref{Fig4}c compares $\eta$ 
across various reported vdW platforms.  
Highlighted in red, our twisted NbSe$_2$ devices strike a 
balance between the rectification strength and critical current.

We next discuss potential ramifications that these nonreciprocal elements can bring to 
quantum circuits and superconducting qubits. For quantum-circuit integration there are multiple standard mitigation routes: (i) operating in a field-free diode regime (several vdW Josephson diodes exhibit $\eta \neq 0$ at $\mathbf{B}\simeq 0$~\cite{wu2022field}); (ii) using \emph{local} on-chip bias lines/coils that generate the required flux only at the diode element while the qubit/resonator is shielded; (iii) using field-compatible qubit materials and vortex-pinning designs (e.g. flux-pinning holes) that allow coherent operation under mT-scale fields~\cite{schneider2019transmon, kringhoj2021magnetic, wang2014measurement}; and
(iv) using innovative qubit designs that are compatible with even larger fields $\sim 0.1\,$T~\cite{Krause2024:PRA}. 
Considering (i)-(iv) allows us to focus directly on the role
of $\eta$, without separately exploring the effect of a global small $\mathbf{B}$.
We model the diode element by an effective current-
phase relation (CPR) $I(\varphi)=I_{\rm C} f(\varphi)$ and define the associated washboard potential $U(\varphi)=E_{\rm J} F(\varphi)$, such that $f(\varphi)=\partial_\varphi F(\varphi)$ and $E_{\rm J}$ is the Josephson energy~\cite{Tafuri:2019}. The anomalous phase $\varphi_0$ describes the situation when $U(\varphi)$ attains the minimum at $\varphi=\varphi_0\neq 0, \pi$, such that $I(\varphi_0)=0$~\cite{Amundsen2024:RMP}, where a simple harmonic form of CPR is given by $f(\varphi)=\sin(\varphi+\varphi_0)$~\cite{Tafuri:2019}. While $\varphi_0$ is symmetry-allowed in the CPR, when inversion and  time-reversal symmetries $\mathcal{P}$ and $\mathcal{T}$ are broken~\cite{Amundsen2024:RMP,Tafuri:2019}, 
it is irrelevant in the context of flux qubit. 
$\varphi_0$ 
just rigidly shifts $I(\varphi)$ and $U(\varphi)$, but has no direct impact on the diode nonreciprocity~\cite{Amundsen2024:RMP}. What matters is that $I(\varphi)$ becomes asymmetric due to a superposition of several (mutually shifted) harmonics, such that $\Delta I=|\max_\varphi I(\varphi)|-|\min_\varphi I(\varphi)|\neq 0$. Here we adopt a minimal parametrization, $f(\varphi)=\sin{(\varphi-\arcsin{\eta})}+\eta$, that directly links $\eta$ with $\Delta I$ following the underlying CPR $I(\varphi)=I_{\rm C} f(\varphi)$; in general such a relation is highly nontrivial. The linear term in $F(\varphi)$ plays the same role as the standard ``tilt'' term in the washboard potential~\cite{Tafuri:2019}
and is used here as an effective description of the diode asymmetry at the circuit level (i.e. not as a statement about microscopic $2\pi$-periodic Josephson energy).
The circuit quantum electrodynamics 
Hamiltonian, $\hat{H}$, for the capacitively shunted Josephson junction is~\cite{Blais-RevModPhys-2021}:
\begin{equation}
\hat{H} = 4E_{\rm C} (\hat{ n} - n_g)^2 + E_{\rm J} F\left(\hat{\varphi}\right),
\end{equation}
where $E_{\rm C}=e^2/(2C)$ and $E_{\rm J}=\Phi_0 I_{\rm C}/2\pi$ are, respectively, 
the charging (capacitative) and Josephson (inductive) energies, with 
$C$ the relevant circuit capacitance, 
 $\Phi_0=h/2e$ the flux quantum, and $n_g$ a gate-induced charge offset~\cite{Tafuri:2019}, 
 while our generalized $F(\varphi)$ deviates from the standard $\cos(\varphi)$ form. 
 Introducing 
bosonic lowering and raising operators 
$\hat{a}$ and $\hat{a}^\dagger$ for an addition and removal of the Cooper pair, the canonically 
conjugated phase and number operators are 
$\hat{\varphi}=\left(2E_{\rm C}/E_{\rm J}\right)^{1/4}(\hat{a}^\dagger + \hat{a})$
and
$\hat{n}=(i/2)\left(E_{\rm J}/2E_{\rm C}\right)^{1/4}(\hat{a}^\dagger - \hat{a})$, 
with 
$[\hat{\varphi},\hat{n}]=i$~\cite{Tafuri:2019}.  
For simplicity we assume that both junctions $J_1$ and $J_2$ are identical and governed by
the minimally skew-symmetric CPR with 
$f(\varphi)=\sin{(\varphi-\arcsin{\eta})}+\eta$, or correspondingly, 
$F(\varphi)=-\cos{(\varphi-\arcsin{\eta})}+\eta\varphi$, where $\eta$ is directly the diode efficiency. 

To proceed further, 
we recall that a superconducting circuit needs a nonlinear element to 
function as a qubit, while for isolating a controllable two-level system 
it is important to have an inherent anharmonicity. We focus on a commercially
successful
transmon qubit with  
$E_{\rm J}\gg E_{\rm C}$~\cite{Tafuri:2019,Blais-RevModPhys-2021}, where the Josephson tunneling dominates, 
the phase-variable takes a well-defined and minimally uncertain value 
$\langle \hat{\varphi}\rangle$ while, desirably, the resulting energy levels
are still largely insensitive to fluctuations in $n_g$, but the cost 
of this protection from the charge noise is the reduced 
anharmonicity~\cite{Tafuri:2019}. For common transmons, denoting by $0, 1, 2$ its lowest states, the qubit transition frequency and the anharmonicity are, respectively,  $\omega_{01}=(\sqrt{8E_{\rm J} E_{\rm C}}-E_{\rm C})$ and 
$\omega_{12}-\omega_{01}=-E_{\rm C}/\hbar$~\cite{Tafuri:2019}.
This analysis of common transmons can be generalized for our case 
where
one expands the underlying goniometric functions entering $F(\varphi)$ around $\langle\hat{\varphi}\rangle$ and gets effective qubit Hamiltonian, $\hat{H}$, in the truncated space; in the case of our CPR this leads to $\langle \hat{\varphi}\rangle\equiv 0$ 
and the truncated Hamiltonian $\hat{H}$ is 
\begin{equation} 
\hat{H} \simeq 4E_{\rm C} (\hat{n}-n_g)^2 
+ 
E_{\rm J}
\left[
\frac{1}{2!} \sqrt{1 - \eta^2} \varphi^2 + \frac{1}{3!} \eta \varphi^3 - \frac{1}{4!} \sqrt{1 - \eta^2} \varphi^4
\right]\,.
\end{equation}
In the above expression we have disregarded numerical constants that are just globally shifting 
the origin of energy. 
The transmon Hamiltonian explicitly breaks 
$\mathcal{T}$ and $\mathcal{P}$ and, therefore, 
the dynamics of the Cooper pairs displays nonreciprocal features.
QuTiP is employed to numerically simulate the potential energy landscape of a transmon diode as a function of phase difference under varying diode efficiency. 
As shown in Fig.~\ref{Fig4}d, the two-level qubit states---the ground ``0" and first excited ``1"---in the tilted cosine-like potential are identified and characterized through this simulation.
The effective potential exhibits a rich spectrum and wide tunability with 
$\varphi$ and $\eta$. A narrow operational regime supporting exactly two quantum states within the central potential well can be created through a precise tuning of 
$\eta$. For this analysis, the careful parameter optimization,  
$\eta \approx 27.6\%$ and $E_{\rm J}/E_{\rm C} = 20.0$, is set to maintain consistency with the experimental observations. 
We find that an intermediate value of $\eta$, not 100\%, remains also optimal for other ratios $E_{\rm J}/E_{\rm C}$.
Typically, for a larger $E_{\rm J}/E_{\rm C}$ a larger $\eta$ would lead to the optimal two-level system configuration.
Additional analysis on how $\eta$ value 
can affect the system is discussed in the Supplementary Information. 
The resulting potential landscape, depicted by the black curve in Fig.~\ref{Fig4}d,
reveals a characteristic asymmetric multi-well configuration that arises from the complex interplay between the Josephson coupling energy and the phase bias introduced by the diode element. Figure~\ref{Fig4}d displays three distinct potential profiles corresponding to different diode efficiencies: $\eta = 10\%$ (left), $\eta = 27.6\%$ (center), and $\eta = 50\%$ (right), each spanning a phase range from $-2\pi$ to $2\pi$. The quantized energy eigenstates are represented as discrete horizontal lines overlaid on each potential well, with the two lowest-energy bound states highlighted in red and blue. These 
states remain spatially localized within the central potential valley and are energetically separated from higher-lying continuum states by the surrounding potential barriers. This evolution
demonstrates how different 
$\eta$ values systematically modify the energy states: $\eta = 10\%$ supports three bound states in the potential well, $\eta = 27.6\%$ maintains exactly two bound states, while $\eta = 50\%$ has only one bound state. The range of $\eta$ that is supporting only two-level configuration is found to be particularly advantageous for quantum computing implementations as it suppresses transitions to higher excited states. 
This can be beneficial for extending 
the qubit coherence time and, generally, for 
functional qubit architectures.

To quantitatively assess the performance of the quantum diode, we introduce the concept of fidelity as a direct measure of transport outcomes. The forward (reverse) fidelity quantifies the accuracy with which an excitation is ultimately transferred in the forward (reverse) direction.
Figure~\ref{Fig4}e compares the forward and reverse transport fidelity across the parameter space of 
$\eta$ and evolution time, with the upper panels showing ideal noiseless evolution and the lower panels incorporating comprehensive noise models. 
There is an optimal $\eta$ for fidelity as its increase has competing effects. A larger $\eta$: (i) enhances the CPR asymmetry, leading to improved directionality and stronger nonreciprocal transport, and (ii) by tilting $U(\varphi)$ reduces the effective potential well and suppresses desirable anharmonicity.
The fidelity between the two states is calculated using the equation 
${\cal F}(|\psi\rangle, |\phi\rangle) = |\langle \psi | \phi \rangle|^2 $, where $|\psi\rangle$ and $|\phi\rangle$ represent the two quantum states of interest. The noiseless simulations (upper panels), the transport fidelities exhibit sharp, well-defined features with maximum forward fidelity approaching unity and minimum reverse fidelity near zero, creating clear high-contrast regions that demonstrate ideal diode behavior with perfect directional selectivity. The noise implementation includes multiple  
relevant sources: Parameter 
fluctuations of $\pm$0.5 in 
$E_J$, charge noise with amplitude 0.03 representing Cooper pair number fluctuations, phase noise of amplitude 0.02 from flux variations, energy dissipation at rate 0.01 modeling environmental coupling, and measurement uncertainty of $\pm$0.002. Despite these realistic noise sources, the characteristic transport asymmetry remains clearly visible, as demonstrated in Fig.~\ref{Fig4}f which shows the overall fidelity difference between forward and reverse directions. The noise primarily introduces broadening and a slight reduction in the peak fidelity values but preserves the fundamental diode behavior, with maximum asymmetry in two levels occurring around $\eta \approx$ 22.9\%-44.8\% and 67.1\%-77.4\% for evolution times of 1-10 ns (see Supplementary Information). 
The comparison validates that the quantum diode effect is sufficiently robust for practical quantum circuit applications, maintaining directional transport characteristics even under typical experimental noise levels, where the signal-to-noise ratio exceeds 10:1 in the optimal parameter regime.

 \section*{Discussion}
In contrast to the current push to maximize the diode efficiency in superconducting diodes and their quantum counterparts, we reveal a much richer picture for optimizing their performance and obtaining functionalities that are not limited to the usual rectifying behavior.
Instead of the efforts to design materials and systems which could attain
a much higher {\em on}/{\em off} ratios and $\eta \rightarrow 100$\%, we put forth a different approach in which already experimentally attained diode efficiency could offer a superior performance of superconducting 
quantum circuits. 
Surprisingly, our findings reveal that even a moderate diode efficiency of 27.6\% in twisted quantum diodes 
is already sufficient to stabilize a well-defined two-level system, a fundamental requirement for reliable qubit operation. This level of nonreciprocity enables a high-fidelity quantum state transport by effectively suppressing backscattering and decoherence pathways. A functional quantum diode could also be designed where the
qubit states define discrete quantized conduction states, introducing 
unexplored functionalities for quantum information storage and processing. The anharmonicity in this quantum diode design is particularly well suited for integration with transmon-based quantum circuits \cite{liu2025strongly}, characterized by $E_{\rm J}/E_{\rm C}\gg 1$ and weakly anharmonic energy levels to prevent unwanted transitions between higher excited states during quantum operations~\cite{Tafuri:2019,Blais-RevModPhys-2021}. 
We show that in our twisted vdW heterostructures a large tunability of the key parameters determines the energy landscape and transform the qubit's properties to overcome the usual reduction of anharmonicity by introducing controlled anharmonicity in the quantum diode. 
This compatibility with transmons can enhance their coherence and control fidelity, making the diode a favorable component for hybrid quantum systems. Such a device could advance quantum computing by enabling precise and energy-efficient data retention. However, the studied rectifying properties may also improve operation of other
superconducting qubits, beyond transmons.
To realize this vision, further experimental research is necessary to fabricate and test various 
quantum diodes and their tunability, including a tantalizing prospect that the resulting Josephson junctions can be turned {\em on} and {\em off} in the absence of any bias current, while the resulting anharmonicity could be desirable for neuromorphic computing~\cite{Monroe2024:APL}. The next critical step will be the integration of these diodes into practical quantum circuits, unlocking many possibilities in quantum-enhanced electronics and computation. 

Given the complexity of materials for quantum diodes and a growing class of candidates that are building blocks for the resulting vdW 
heterostructures, machine learning techniques are being explored to accelerate their materials discovery and interface 
optimization~\cite{leger2024machine,zhang20242d,sosa2025simulating}. Using AI-driven approaches 
could predict and design materials combinations to minimize losses and enhance efficiency, making quantum diodes
viable for 
energy-efficient and noise-resilient quantum circuits. Conversely, a similar approach would also enable exploring a large phase space of materials parameters and the discovery of fundamental properties in twisted superconducting diodes. Given that diode effects 
can characterize Josephson junctions~\cite{Pekerten2024:APL} and accompany topological superconductivity~\cite{Amundsen2024:RMP}, there is an intriguing prospect of using diode effects in exploring topological superconductivity in twisted vdW systems.

\section*{Methods}
{\large Device fabrication}

Pristine NbSe$_2$ are commercially bought from a company. The thin films are obtained from mechanical exfoliation using conventional scotch tape. The metal electrodes are prepared using photolithography, followed by e-beam evaporation of 5 nm/30 nm Cr/Au films. The heterostructures were assembled using transfer stage, with assistance of PDMS/PPC polymers to hold the layers. After the assembly of the heterostructure, all layers were transferred onto the prepared metal electrodes, then using acetone to remove polymer layers. 

{\large Transport measurement}

Four-probe geometry current vs voltage scans were measured to observe the critical current nonreciprocity. The measurement was done in our Physical Property Measurement System (PPMS) with a temperature range of 1.6 K-400 K and a magnetic field of $\pm$9 T. A DC source is applied from Keithley 2450 and Keysight B2902A.

{\large Degauss procedure}

Before each measurement of the superconducting diode effect, the superconducting magnet was reset to minimize the influence of trapped magnetic flux. The magnet was first ramped to $100~\mathrm{mT}$ and then returned to zero field through several oscillatory sweeps with gradually decreasing amplitude. The field ramp rate during this procedure was approximately $10~\mathrm{mT/min}$.

{\large Sample alignment with magnetic field}

The sample was mounted on a PPMS-compatible rotatable probe. During mounting, the source--drain direction of the device was first manually aligned with the nominal magnetic-field direction. To calibrate the actual sample-field alignment, we performed full $360^\circ$ rotation-dependent transport measurements at $10~\mathrm{K}$ in both the in-plane and out-of-plane rotation geometries. The angular dependence of the resistance was used to determine the alignment offset: the resistance maximum in the out-of-plane rotation and the resistance minimum in the in-plane rotation identify the corresponding $0^\circ$ reference direction. The extracted angular offset was then used to define the field-alignment condition for subsequent measurements, accounting for possible mounting tilt and mechanical hysteresis of the rotator.

{\large QuTip quantum simulation}

The quantum simulation presented in this study utilizes the Quantum Toolbox in Python (QuTiP), a robust open-source software designed specifically for quantum mechanics and quantum computing simulations. QuTiP allows efficient numerical calculations involving quantum states, density matrices, Hamiltonians, and time evolutions, making it 
particularly suitable for modeling complex quantum systems such as the multi-qubit quantum diode discussed herein. In this simulation, QuTiP facilitates the representation and manipulation of quantum states with Josephson junction providing the diode efficiency, the implementation of time evolution operators, and the inclusion of realistic quantum noise and decoherence effects. By leveraging QuTiP's powerful 
computational tools, the forward and backward quantum state transfer probabilities were
systematically 
explored under varying levels of coupling asymmetry and noise conditions, accurately capturing the dynamics essential for analyzing quantum diode performance.

\section*{Data Availability}
The data supporting the findings of this study are publicly available on Zenodo \cite{Zhong2026SuperconductingDiode}.

\section*{Code Availability}
The code used to generate and analyze the results presented in this study is publicly available on Zenodo \cite{Zhong2026SuperconductingDiode}.

\printbibliography

\section*{Acknowledgments}
We want to thank Dr.~Yasen Hou for help in the experiment setup and meaningful discussions. Support from UF Gatorade award and Research Opportunity Seed Fund are kindly acknowledged. This material is also based upon work supported by the National Science Foundation under Grant No. 2441051 and No. 2501208 (Y. Wu). Partial support is acknowledged from the Slovak Academy of Sciences (IMPULZ 2021 project IM-2021-26 “SUPERSPIN”), the Slovak Research and Development Agency (contract No.~APVV-24-0134), and the National Science and Technology Council of Taiwan (Grant No.~114-2112-M-006-034-MY3) (D.~Kochan). This work is supported by the U.S. Department of Energy, Office of Science, Basic Energy Sciences under Award No. DE-SC0004890 (I.~\v{Z}uti\'c). 

\section*{Funding statements}
This work was supported by the University of Florida Gatorade Award and Research Opportunity Seed Fund, and by the National Science Foundation under Grant Nos. 2441051 and 2501208 (Y.Wu). D.Kochan acknowledges support from the Slovak Academy of Sciences through the IMPULZ 2021 project IM-2021-26 “SUPERSPIN,” the Slovak Research and Development Agency under Contract No. APVV-24-0134, and the National Science and Technology Council of Taiwan under Grant No. 114-2112-M-006-034-MY3. I.~\v{Z}uti\'c acknowledges support from the U.S. Department of Energy, Office of Science, Basic Energy Sciences under Award No. DE-SC0004890.

\section*{Author Contributions}
Y.~Wu conceived the ideas, supervised the project and led
with I.~\v{Z}uti\'c the manuscript writing. 
H.~Zhong led sample fabrication, experiment measurements, data collection and quantum simulations. D.~Kochan proposed truncated qubit Hamiltonian,
together with I.~\v{Z}uti\'c they interpreted results and gave theoretical guidance. 
All authors discussed and commented on the manuscript and also contributed to the manuscript revision. 

\section*{Competing Interests}
The authors declare no other competing interests.

\end{document}